# APPLICATION OF ROBUST ESTIMATORS IN SHEWHART S-CHARTS

**Burak ALAKENT***
Boğaziçi University, Chemical Engineering Department
**İstanbul, Turkey.**

**Ece Çiğdem MUTLU**
Boğaziçi University, Chemical Engineering Department
**İstanbul, Turkey.**

** Corresponding author: Burak Alakent, Phone: 02123596433*
*E-mail address: burak.alakent@boun.edu.tr*

## ABSTRACT

Maintaining the quality of manufactured products at a desired level is known to increase customer satisfaction and profitability. Shewhart control chart is the most widely used in statistical process control (SPC) technique to monitor the quality of products and control process variability. Based on the assumption of independent and normally distributed data sets, sample mean and standard deviation statistics are known to be the most efficient "conventional" estimators to determine the process location and scale, respectively. On the other hand, there is not guarantee that the real-world process data would be normally distributed: outliers may exist, and/or sampled population may be contaminated. In such cases, efficiency of the conventional estimators is significantly reduced, and power of the Shewhart charts may be undesirably low, e.g. occasional outliers in the rational subgroups (Phase I dataset) may drastically affect the sample mean and standard deviation, resulting a serious delay in detection of inferior products (Phase II procedure). For more efficient analyses, it is required to use "robust" estimators against contaminations, which may exist in Phase I. Here, we present a simple approach to construct robust control charts using M-Huber, Harrell Davis and Hodge Lehmann estimators for monitoring the process location, and logistic M scale (MSLOG), median absolute deviation (MAD) and $Q_n$ estimator of Rousseuw & Croux (QN) for monitoring process scale. Performance of various robust estimators are compared with the conventional mean and standard deviation statistics, in terms of their Phase I efficiencies and Phase II average run lengths (ARLs). A Monte Carlo simulation study is conducted via MATLAB to compute the required statistics and performance criterion. Consequently, it is determined that robust estimators are more efficient both against diffuse-localized and symmetric-asymmetric contaminations, and have higher power in detecting disturbances, compared to conventional methods.

## INTRODUCTION

Stability of the scale and location parameters of a process, and prompt detection of perturbations in these parameters play significant roles in process efficiency. Statistical process control (SPC) is a group of tools and techniques to monitor the quality of products and control process variability. Monitoring can be performed via check sheets, cause and effect sheets, flow charts, pareto charts, scatter diagrams or histograms, all of which give a summary of a process as discrete snapshots. Control charts, on the other hand, enable to dissect the variation during time period that the process is observed. This is why control charts are widely used techniques to monitor quality of products.

Control chart was first proposed in 1924 by Dr. Walter Shewhart to analyze the variability in a process and to detect the disturbances (Shewhart, 1931). In the following years, numerous studies have been published on application and modification of control charts (Duncan, 1986; Schoonhoven et al., 2012; Montgomery & Runger, 2008). The main purpose of the online application of control charts is to determine unusual changes by eliminating "special (assignable) causes" from "common (natural) causes". This paradigm aims to keep the process under control by preventing potential causes rather to extinguish in the past (Montgomery & Runger, 2008). Common causes, which creates acceptable variability, are caused by the inherent factors in a process. They don't have a prominent effect on equilibrium point. On the other hand, special causes, which usually stem from assignable contaminations, tend to affect the equilibrium point undesirably, and special causes should be detected and eliminated from the process to reduce variability and improve quality (Woodall,1999).

In SPC, process monitoring is mainly handled in two consecutive steps. In Phase I, data are collected as rational subgroups. The main purpose of rational subgrouping is to exclude the variability, which may stem from special causes. Using rational subgroups, variability within subgroups is guaranteed to represent the natural variability dictated by common causes. Rational subgrouping is considered as an essential step for process monitoring, and Nelson (1988), Sefik (1998) and Montgomery (2008) demonstrated proper procedures of rational subgrouping. Using the data in the subgroups collected from the in-control process, table process parameters are estimated retrospectively. Using these estimated parameters, upper and lower control limits (CLs) are determined for Phase II, in which hypothesis tests are continuously performed for new observations from the process (Montgomery & Runger, 2008).

In the current work, standard deviation (S) control charts are exclusively studied. A typical Shewhart S-control chart (Figure 1) is composed of a central line (CL), which represents the expected value of the quality variable under common causes, and upper-lower control limits. These control limits are determined so that 99.73% of all sample points fall between CLs if the process is in control. Process is deemed to be out of control if sample points fall beyond these limits and, causes responsible the change should be investigated and eliminated.

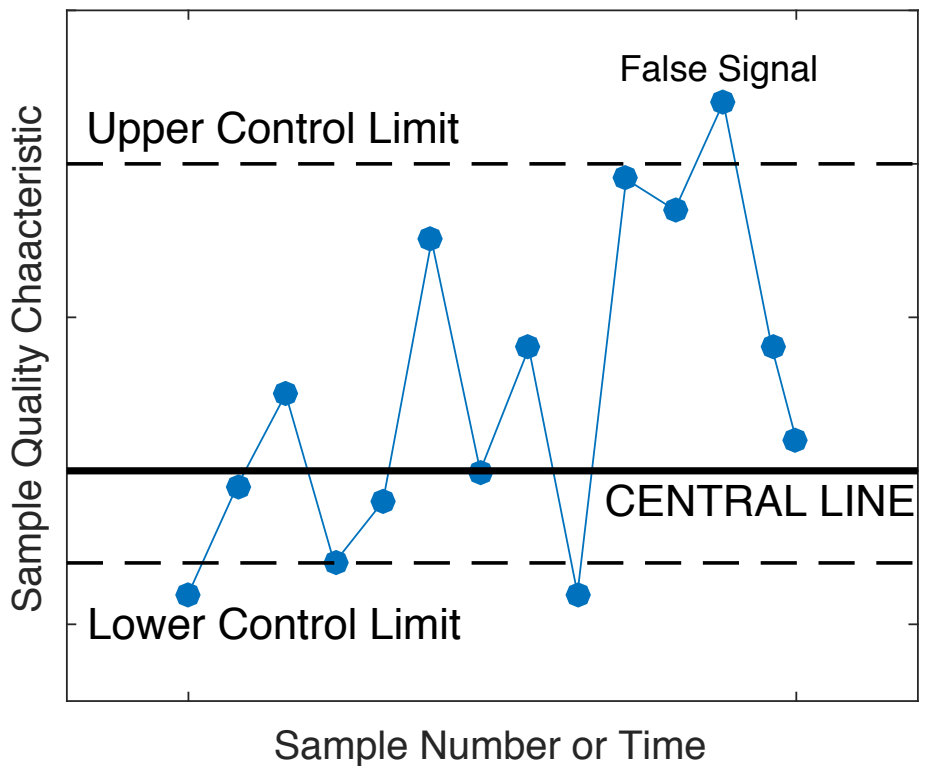

**Figure 1:** A typical Shewhart S-control chart

In the design of S control chart for a process assumed to be in control, scale parameter of each rational subgroup is estimated ($\hat{\sigma}_i{}^{I}$) during Phase I. Denoting the sample and subgroup sizes of Phase I data by n and k, respectively, $X_{ij}$, ($i = 1,2,\dots,k$ and $j = 1,2,\dots,n$) represents individual observations, assumed to be independent and normally distributed $N(\mu,\sigma^2)$ distributed. Phase II data consist of an indefinite number of samples with the same sample size, and each new observation, denoted with $Y_{ij}$, ($i = 1,2,\dots$ and $j = 1,2,\dots,n$), is assumed to be independent and $N(\mu,(\phi\sigma)^2)$ distributed. Here, $\phi = 1$ represents an in-control process, while $\phi \neq 1$ corresponds to a process under a disturbance. In a S control chart, central line is taken to be equal to the location of scale parameters of subgroups with convenient upper and lower control limits, and the standard deviation of each subgroup ($\hat{\sigma}_i{}^{II}$) in Phase II is placed in this chart successively.

It should be noted that the standard deviation of the process is unknown, so the Phase I estimate ($\hat{\sigma}$) is substituted for the real standard deviation ($\sigma$) of the process in determining the lower control limit (LCL) and upper control limit (UCL) estimates

$$\widehat{LCL} = L_n\hat{\sigma}, \qquad \widehat{UCL} = U_n\hat{\sigma} \tag{1}$$

Here, $L_n$ and $U_n$ are chosen so as to obtain the desired type I probability in Phase II sampling.

$$P\left(L_n\hat{\sigma} \leq \hat{\sigma}_i{}^{II} \leq U_n\hat{\sigma}\right) = 1 - \alpha \tag{2}$$

Furthermore, if the process is exposed to a disturbance ($\phi \neq 1$) in Phase II, out of control data should be detected as soon as possible. We define $F_i$ as the event that the standard deviation estimate of $i^{th}$ subgroup in Phase II ($\hat{\sigma}_i{}^{II}$) falls beyond the CLs determined in Phase I, and $P(F_i) = p$ is the probability of the occurrence of that event. Run length (RL) is the number of subgroups, which have not given alarm between two consecutive out-of-control signals. If standard deviation of the process is known, RL has a geometric distribution (Montgomery & Runger, 2008). In this case, Phase I procedure may be skipped. In real world processes, however, standard deviation is usually unknown and required to be estimated. In such cases, the probability distribution of RL is not geometric and its expected value may be determined using Monte Carlo (MC) simulations. Each Phase I MC sampling yields a different conditional run length the unconditional average run length (ARL) may be computed using the average of a large number of conditional run lengths.

$$ARL = E\left\{E(RL|\hat{\sigma})\right\} = E(RL) = E\left(\frac{1}{p}\right) \cong \sum_{i=1}^{M}\frac{1}{p_i} \tag{4}$$

Taking the limits enclosing 99.73% of normal distribution to be acceptable to deem a process to be in control, it is common practice to use the unconditional $ARL_0$ (ARL for a in-control process) to be equal to 370.4. This means that even if the process is in control, a false signal is expected in every 370 samples (Schoonhoven & Does, 2012).

The rest of study is structured as follows. After a brief summary on robust estimators is given, different types of disturbances (contaminations), which cause temporary outliers in the Phase I data of the process, are introduced. The proposed location and scale estimators with their robustness properties are presented in the next section. Results and Discussion will consist of efficiencies Phase I estimators, $ARL_0$ and ARL in Phase sampling. A brief conclusion section will summarize the findings.

## METHOD

In statistical process control, sample mean and sample standard deviation statistics are used to determine scale and location of the process based on the assumption of having independent and normally distributed dataset. On the other hand, the real world data may not be normally distributed due to miscalibrated equipment, operator errors and/or raw materials coming from different resources; outliers may exist and/or sampled population may be contaminated (Tatum, 1997). The conventional estimators are known to be most efficient and powerful statistics when employed on normally distributed data, but existence outlier(s) in a dataset may perturb the sample mean and standard deviation drastically. In such cases, robust estimators, which are less sensitive to the outliers, may be used to resist outliers. Breakdown point (BDP) and influence function (IF) are the most common measures of robustness. Asymptotic BDP measures the percentage of contamination (practically, outliers in a sample), which can render the functional form of the estimator meaningless in a population. IF, on the other hand, is used to measure the effect of a single outlier on the estimator. Hence, BDP and IF are used to measure the "global" and "local" robustness of an estimator, respectively. A robust estimator should have a bounded influence function (BIF) and high breakdown point, to resist against outliers (Rousseeuw & Verboven, 2002).

### Outlier Models in Phase I

In order to compare the Phase I efficiencies of proposed estimators and the power of the control charts constructed using the CLs determined from the proposed estimators, three different contamination models, likely to represent outlier producing mechanisms encountered in real world, are suggested (Tatum,1997).

In the following model, phase I data will be assumed to be coming from a standardized normal variable ($Z_{IC} \sim N(0,1)$) representing the in-control process, coupled with another random variable ($Z_{out}$, from various distributions for different outlier models) representing temporary out-of-control states of the process.

***Model 1:*** *Diffuse Symmetric Contamination*
Each observation has 0.80 probability of being drawn from the population of $Z_{IC}$, and 0.20 probability of being drawn from the population of $Z_{out} \sim N(0, a^2)$ with $a = 2, 3, 4$. As a result, outliers are scattered to the subgroups symmetrically.

***Model 2:*** *Diffuse Asymmetric Contamination*
Each observation has a 0.80 probability of being drawn from the population of $Z_{IC}$, and 0.20 probability of being drawn from the population of $Z_{out} \sim \chi_a^2$ (a denoting the number of degrees of freedom) with $a = 2, 3, 4$. Hence, outliers are scattered to the subgroups asymmetrically.

***Model 3:*** *Localized Contamination*
Randomly chosen 80% of all subgroups are sampled from the population of $Z_{IC}$, while the remaining 20% are sampled from the population of $Z_{out} \sim N(0, a^2)$ with $a = 2, 3, 4$. Here, outliers are confined into certain subgroups, while the rest of the subgroups are free from outliers.

### Proposed Scale Estimators

In this study, M-Huber, Harrell Davis and Hodge Lehmann estimators for monitoring the process location, and logistic M scale (MSLOG), median absolute deviation (MAD) and Qn estimator of Rousseuw & Croux (QN) for monitoring process scale have been used as robust estimators.

In the design of S control charts, rational subgroups are collected and scale parameters of each rational subgroup are estimated, followed by the estimation of the location of subgroup scales. In this section, scale parameter estimators are reviewed, while following section comprises location estimators in the current study.

In practice, conventional statistics are generally used in Phase I to estimate the standard deviation $\sigma$ of in-control process and the standard deviation $\phi\sigma$ in Phase II. This method is liable to give imprecise estimates in case of contaminations of Phase I data (Jensen et al., 2006).

Sample standard deviation is known to be the most efficient conventional scale estimator under normality. If each observation in a sample is represented with $x_i$, ($i$ = 1,2, ... n), then standard deviation estimate is as follows.

$$\widehat{S_n} = \sqrt{\frac{1}{n-1}\sum_{j=1}^{n}(x_n - ave_n(x))^2} \tag{5}$$

Sample standard deviation has 0 % breakdown point, i.e. a single outlier has the potential to change the scale estimate indefinitely. As a result, influence function of sample standard deviation is unbounded.

Median absolute deviation (MAD) is the unbiased median estimate of absolute deviations from median.

$$MAD_n = b_n med_n |x_n - med_n x| \tag{6}$$

$MAD_n$ has 50% breakdown point and bounded influence function, making $MAD_n$ less sensitive to outliers. It is one of the widely used median based estimators due to its good robustness properties and its simplicity. Although $MAD_n$ has low (37%) Gaussian efficiency, it may be highly efficient when the sampled population is contaminated.

Another type of scale estimator named QN estimator was suggested by Rousseuw & Croux (1993).

$$Q_n = c_n med\{|x_i - x_j|; i < j\} \tag{7}$$

In Equations 6 and 7, $b_n$ and $c_n$ are the correction factors which make $MAD_n$ and QN unbiased estimators respectively. While $Q_n$ has also 50% breakdown point and bounded influence function, but, discontinuities in the influence functions of $MAD_n$ and QN make the application of these estimators less favorable in small samples (Rousseuw and Verboven, 2002). The advantage QN over $MAD_n$ is its high Gaussian efficiency (≈83 %).

M logistic scale estimator (MSLOG) is an M-estimator of scale with psi-function equal to $(e^x - 1)/(e^x + 1)$ and with auxiliary location estimate.

$$M_n = \frac{1}{n}\sum_{i=1}^{n}\rho\left(\frac{x_i - \hat{\mu}_0}{\sigma}\right) = \kappa, \quad 0 < \kappa < \rho(\infty) \tag{8}$$

In the above equation, constant $\kappa$ can be adjusted to attain the desired breakdown point. In the current study, a maximum breakdown point 50% is used in MSLOG estimator. Influence function of MSLOG is smooth and bounded. Therefore, fully iterated M estimator with logistic psi function is used to prevent sudden bumps in influence function.

### Proposed Location Estimators

Sample mean (also called the sample average) is the most widely used location estimator. Given a set of observations $x_i$, ($i$ = 1,2, ... n), sample mean is defined as;

$$\bar{x} = \frac{1}{n}\sum_{i=1}^{n} x_i \tag{9}$$

Sample mean is not resistant against disturbances due to low breakdown point (0%) and unbounded influence function. Especially when the sample size is small, sample mean is affected from outliers excessively.

One of robust estimators to resist contaminations is the M-estimator, which is suggested by Huber (Huber,1964).

$$\sum_{i=1}^{n} \varphi\left(\frac{x_i - T}{\sigma}\right) = 0, \quad \varphi(x) = min\,(c, max(x, -c)) \qquad (10)$$

Here, the influence function $\varphi$ is a linear function of the normalized variable $x$ for $|x| < c$, and constant for $|x| > c$, in which c is usually taken to be equal to 1.5 for a satisfactory efficiency of the estimator. An auxiliary scale estimator $\sigma$ is required to determine the location M-estimator $T$. The maximal breakdown point (50%) can be achieved with M-estimator, which has a bounded influence function.

Another well-known robust location estimator is Harrell-Davis quantile estimator (Eqn. 11), which is the weighted average of all order statistics (Harrell & Davis, 1982). One of the direct applications of Harrell-Davis estimators is in estimating the sample median. Compared to sample median, Harrell-Davis estimator is a highly efficient, since it uses all of the observations rather than the order statistics. It should, however, be noted that a single outlier, if sufficiently distant from the rest of the observations, may render this statistic useless. One way to take a precaution against this phenomenon while using Harrell-Davis estimators is to use robust scale estimators used on the subgroups, as performed in the current study.

$$P(Y) = \frac{\Gamma(a+b)}{\Gamma(a)\Gamma(b)} X^{(a-1)}(1-X)^{(b-1)}, \;\; X_i = \sum_{i=1}^{n} W_i X_i \qquad (11)$$

The final location estimator, considered in the current study, is the Hodges-Lehmann estimator which is median of the Walsh averages (Hampel, et al., 1986). Because it is a median based estimator, its influence function is bounded. Despite its robustness, its influence function may coincide with the sample mean when sample size is low. Additionally, its breakdown point is ~29%, which is relatively low.

$$X_i = med\left\{\frac{X_i - X_j}{2}; 1 \le i < j \le n\right\} \qquad (12)$$

## RESULTS AND DISCUSSION

Phase I efficiency of the standard deviation estimators listed above is assessed via computing mean square errors (MSEs). MSE of a standard deviation estimator (Eqn. 13) is equal to the expected value of the squared difference between the estimate ($\hat{\sigma}_i$) and the true standard deviation ($\sigma$). MSE comparisons are performed for the three types of contamination models, which has been explained above.

$$MSE = \frac{1}{N}\sum_{i=1}^{N}(\hat{\sigma}_i - \sigma)^2 \qquad (13)$$

Figures 2-4 show the change of MSEs with respect to the contamination parameter ($a$) for outlier models 1-3, respectively. Each figure consists of four subfigures, in each of which a different location estimator is applied to the scale estimates of the subgroups. Each figure exhibits MSEs of four different scale estimators: sample standard deviation (blue sharp line), $MAD_n$ (dashed line), QN (filled circle marker) and MSLOG (circle marker) estimators for the same location estimator. Hence, a total of 16 different MSE representations correspond to all the combinations of subgroup scale estimators + location estimators applied on Phase I data. It should be noted that $a = 1$ corresponds to no contamination for outlier models 1 and 3. In Figure 2, sample standard deviation, as expected, has the lowest MSE, when Phase I samples are drawn solely from the population of $Z_{IC}$. However, as contamination is increased, e.g. when $a$ exceeds two, efficiency of sample standard deviation is exceeded by the robust scale estimators for all location estimators. For $3<a<4$, efficiency of $MAD_n$ is the highest, closely followed by MSLOG estimator, while efficiency of $MAD_n$ decreases for $a < 2$-2.5. When mean is applied as the location estimator (Figure 2, upper-left subfigure) and $a$ is equal to 4, MSE of sample standard deviation is approximately 0.66, whereas, those of $MAD_n$, MSLOG and QN are equal to 0.17, 0.19 and 0.22, respectively. Standard deviation estimation can further be improved using robust location estimators, instead of sample mean. Here, Harrell-Davis (upper-right) location estimator yields smaller MSEs, followed by M-Huber (lower-left) and Hodge-Lehmann (lower-right) estimators. For the maximum contamination level, MSE values obtained from the $MAD_n$ estimator are found to be equal to 0.17, 0.11, 0.12 and 0.12 for Harrell-Davis, M-Huber and Hodge-Lehmann estimators, respectively. Though, among all combinations, $MAD_n$+Harrel-Davis standard deviation estimator has the highest efficiency as $a$ approaches its upper limit, for moderate values of $a$, its efficiency is exceeded by that of MSLOG based estimators, particularly MSLOG+Harrel-Davis estimator. This shows that MSLOG+Harrel-Davis estimator is convenient for processes, in which a mild level of contamination in the data is expected.

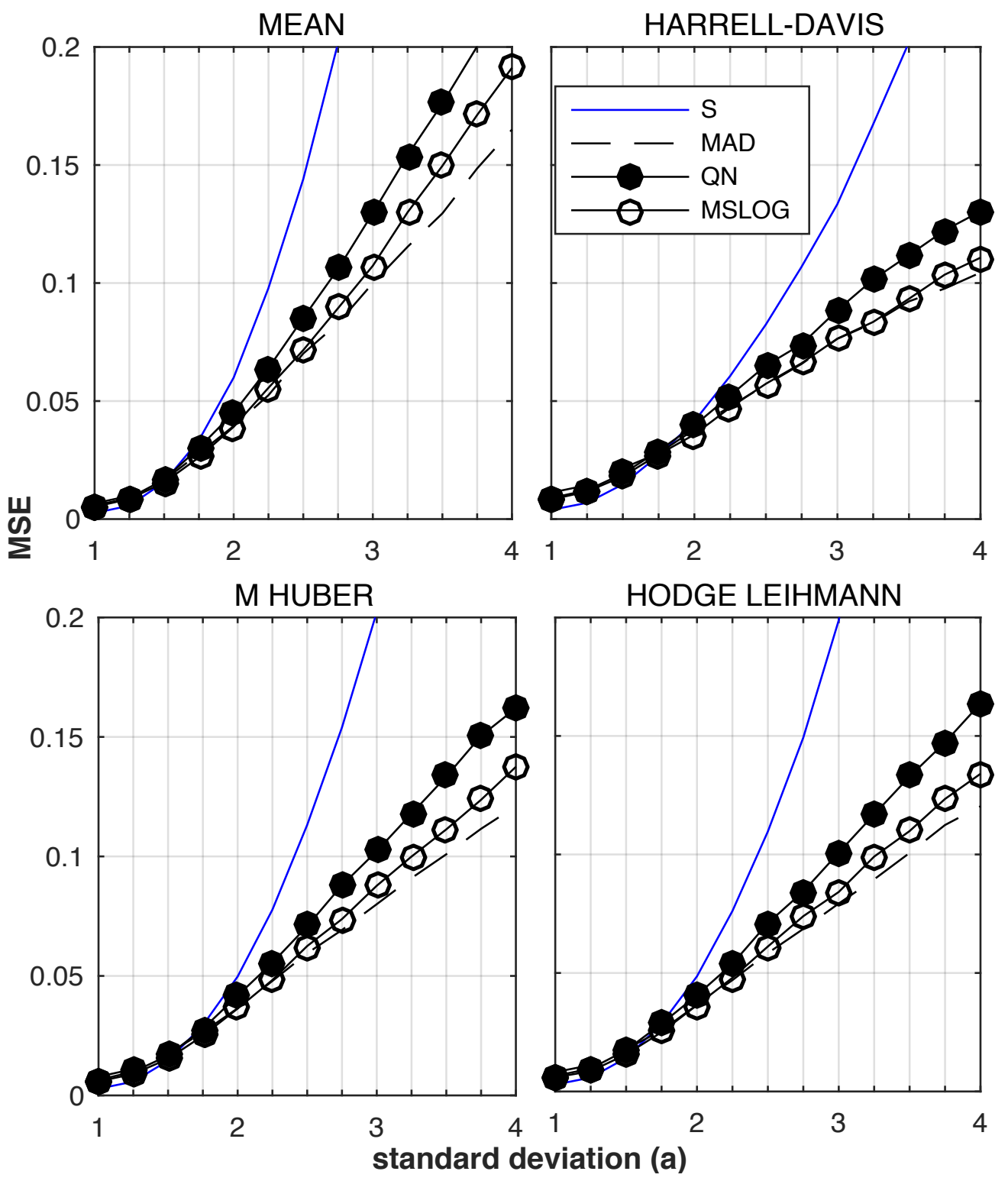

**Figure 2:** MSE of estimators under diffuse symmetric contamination (model 1) in Phase I

Effect of asymmetric diffuse contamination (model 2) on the efficiency of sample standard deviation is even harsher (Figure 3). Here, the smallest MSEs are attained by QN estimator, distantly followed by $MAD_n$ and MSLOG estimators. Among the location estimators, Harrell-Davis estimator, again, seems to be superior compared to M-Huber and Hodge-Lehmann estimators. The estimator combination with the smallest MSE is QN+Hodge-Lehmann estimator, but the superiority of this estimator to the other two robust location estimators is negligible.

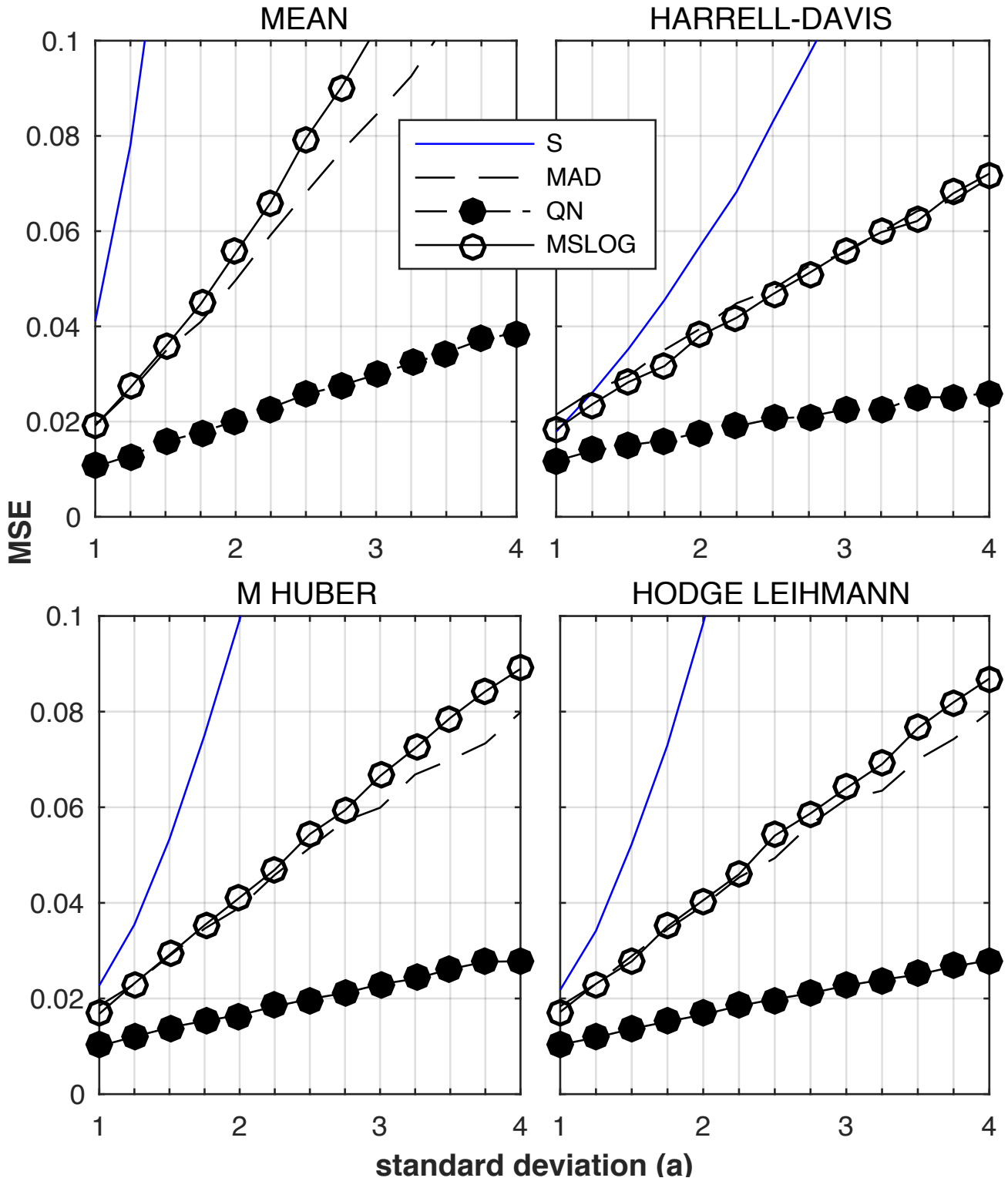

**Figure 3:** MSE of estimators under diffuse asymmetric contamination (model 2) in Phase I

MSEs of the estimators on Phase I data under localized contamination (model 3) may, at first sight, seem surprising, since sample standard deviation, unlike the first two outlier models, has a relatively high efficiency. It should, however, be recalled that, here, there are no outliers in 80% of the subgroups, and the remaining subgroups are composed solely of contaminated (still normal but with a different variance) observations. Hence, one should not expect robust scale estimators of subgroups to handle the outlier problem, but robust location estimators should be effective. Since sample standard deviation is the most efficient scale estimator when sampled data is normally distributed, its MSE is lower than MSLOG, QN and $MAD_n$ estimators. Among the location estimators, Harrell-Davis estimator yields the smallest MSEs if $a > 2$. For smaller contaminations, performance of Harrell-Davis estimator lags slightly behind those of M-Huber and Hodge-Lehmann estimators. As a result, one may say that sample standard deviation+Hodge-Lehmann estimator seems to be a proper choice for point estimation of standard deviation of processes with moderate level contamination in its Phase I data.

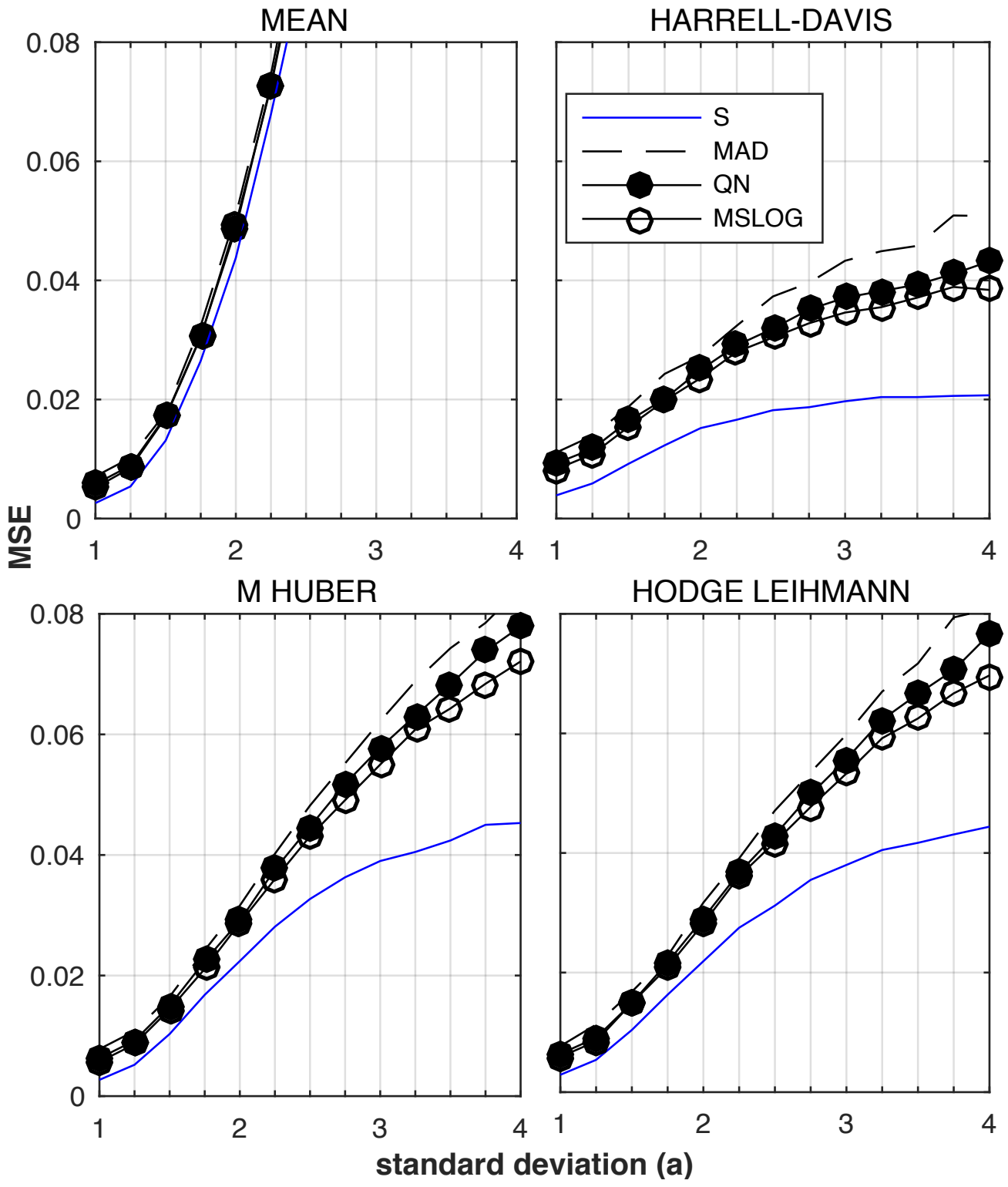

**Figure 4:** MSE of estimators in the existence of localized contamination (model 3) in Phase I

To be able to compare Phase II performances of the suggested estimators, Phase II CLs of each estimator is constructed such that unconditional $ARL_0$ is equal to 370.4 under normal distributed (no contamination present) Phase I data. Phase II performance is assessed using two criteria. First, even though Phase I data may contain outliers, Phase II data will be collected from the in-control process, so $ARL_0$ of the estimator in question should be as high as possible with respect to the design value 370.4. Here, it is aimed to reduce the false positive signals. Second, Phase II samples in the out-of-control state of the process should be detected as soon as possible, i.e. ARL should be low. The criterion is related with the power of the control chart scheme.

Figure 5 shows the $ARL_0$ values for the case when model 1 is applied to Phase I sampling, and Phase II data is normally distributed with $\phi = 1$. When conventional methods are used (blue square marker), $ARL_0$ is 305.6, much lower than the design value even at low contamination, and drastically drops down to 78.1 at high contamination, showing that the conventional statistics would be liable to give false signals. Furthermore, it is seen that both the subgroup scale and location estimators should be robust for a high $ARL_0$ value. Here, the highest $ARL_0$ values are obtained using $MAD_n$ and MSLOG estimators, coupled with any of the robust location estimators.

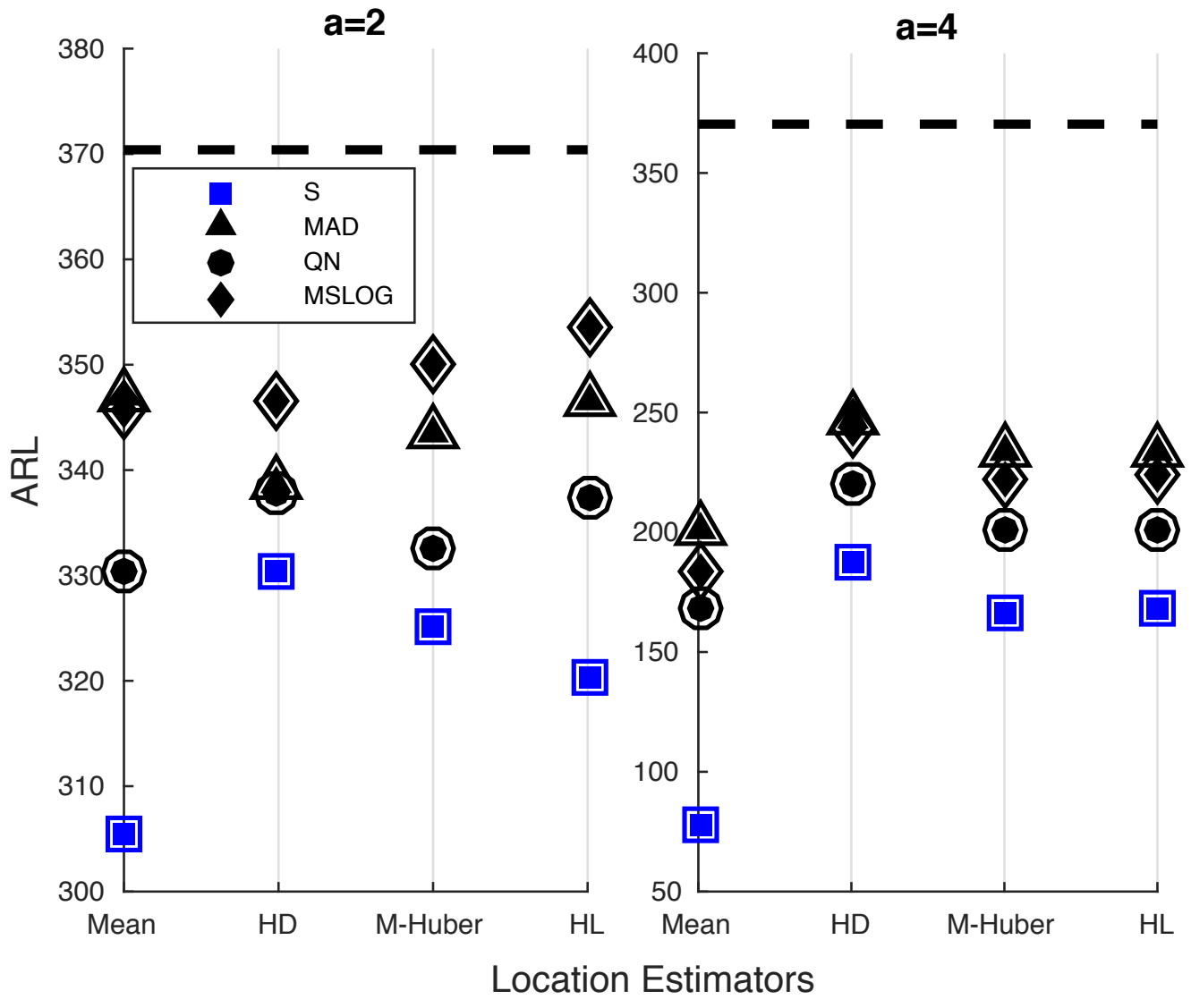

**Figure 5:** In control ARLs in the existence of model 1

Figure 6 shows the ARL values for the case when model 1 is applied to Phase I sampling, and Phase II data is normally distributed with $\phi = 1.4$ (disturbance applied in Phase II). Here, although conventional methods are the most powerful statistics when contamination does not exist (a = 1), as contamination increases, inferior products cannot be detected promptly. For example, at the highest levels of contamination, ARL increases to 480.9 when sample standard deviation and sample mean are used, but robust statistical methods, particularly MSLOG+HD, can detect the disturbance at a relatively lower value of ARL=287.5. At a moderate level of contamination, however, MSLOG+M-Huber or HL estimators seems to be a proper choice for high power.

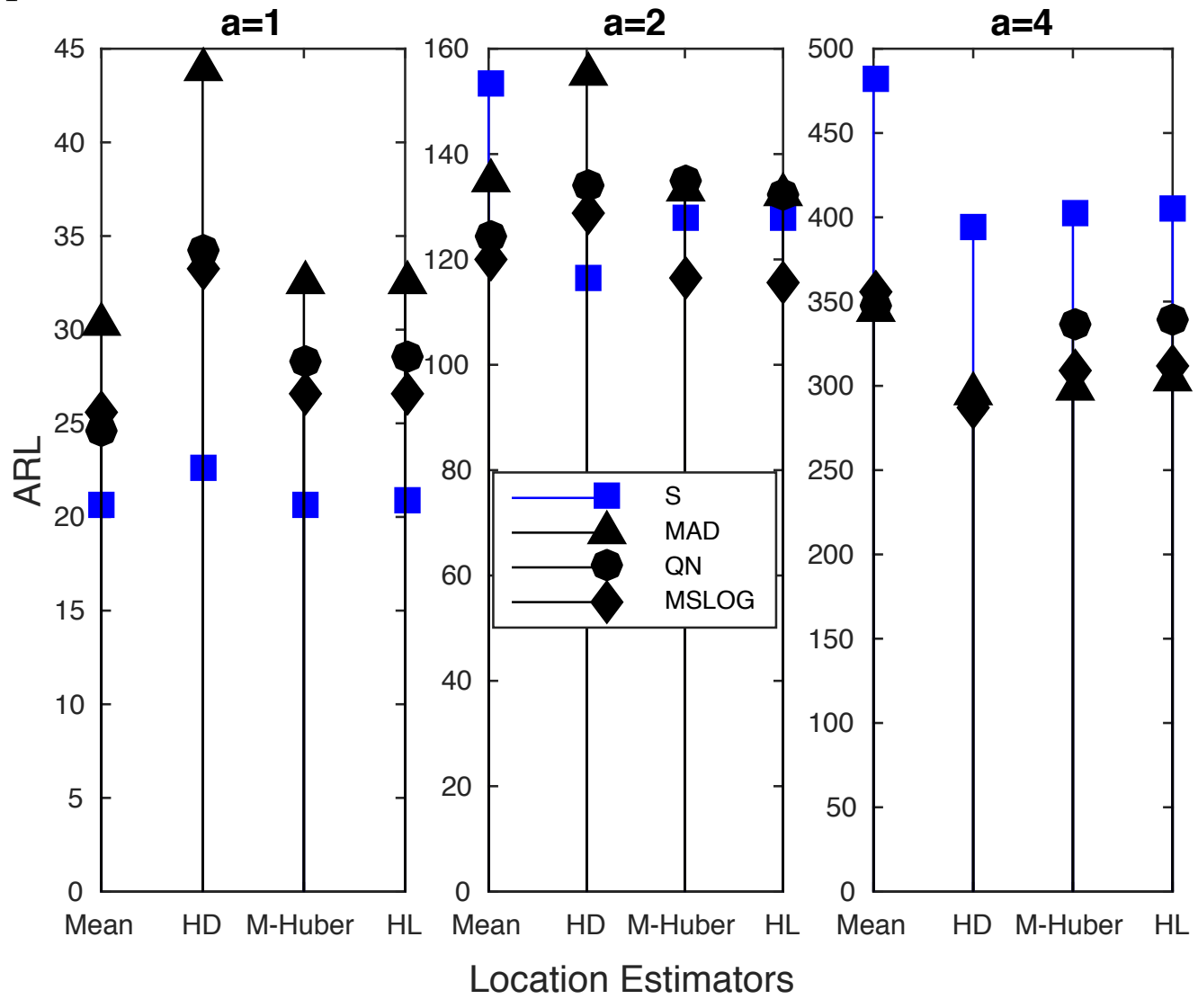

**Figure 6:** Out of control ARLs in the existence of model 1

Figure 7 shows the $ARL_0$ values for the case when model 2 is applied to Phase I sampling, and Phase II data is normally distributed with $\phi = 1$ (no disturbance in Phase II). For highly contaminated Phase I data, conventional methods yield a false at every 38.5 samples on the average, whereas $ARL_0$ for QN is found to be equal to 364, close to the design value. When Phase II sampling is performed from disturbed process ($\phi = 1.4$), QN estimator is the most powerful scale estimator whether the size of contamination is low or high (Figure 8). At moderate and high contamination level, QN+Huber-M seems the best choice in terms of $ARL_0$ and ARL values. The conventional sample standard deviation statistic can be seen to be more powerful than robust scale statistics at low contamination, its specificity and power are decreased with increasing size of contamination.

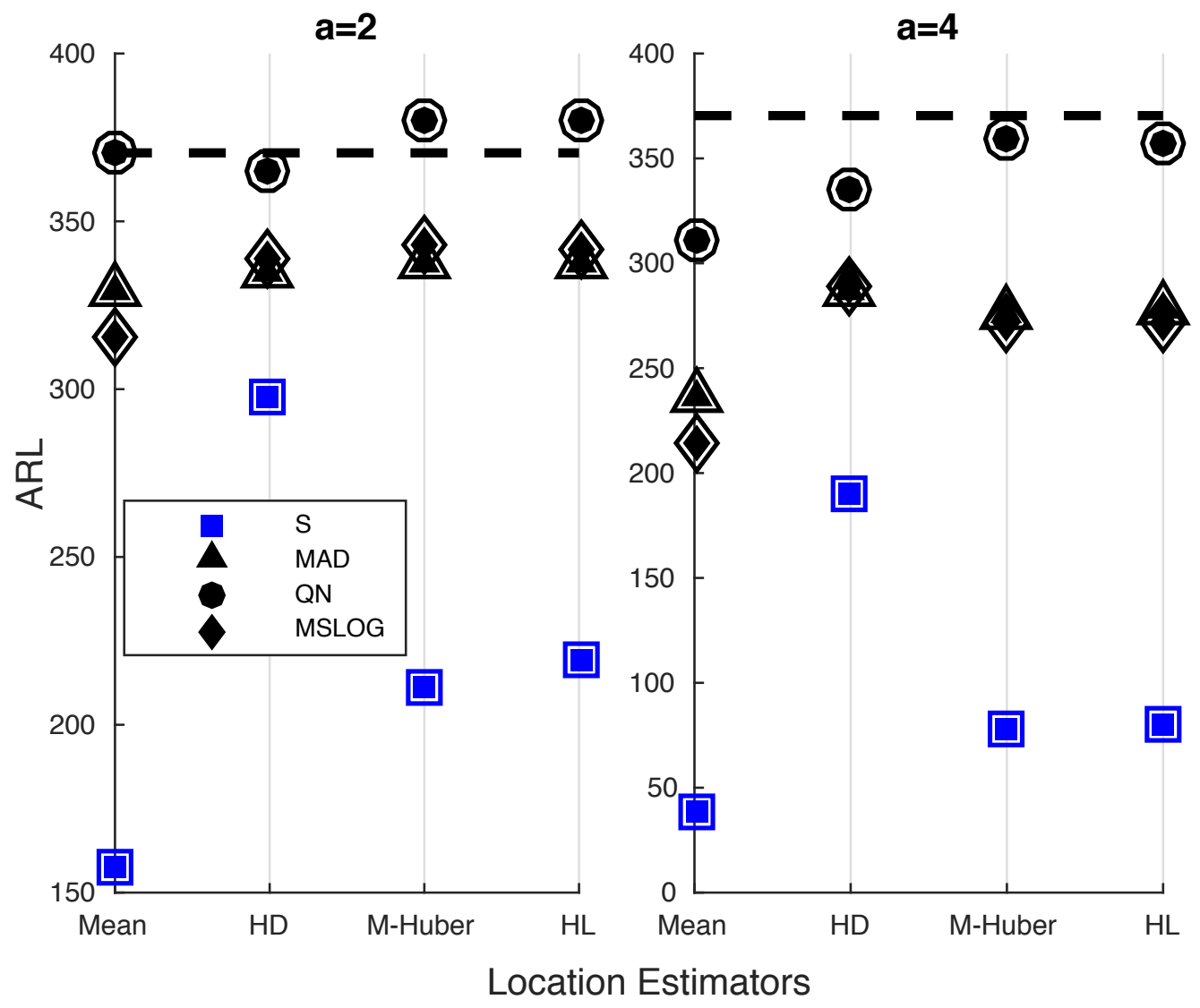

**Figure 7:** In control ARLs in the existence of model 2

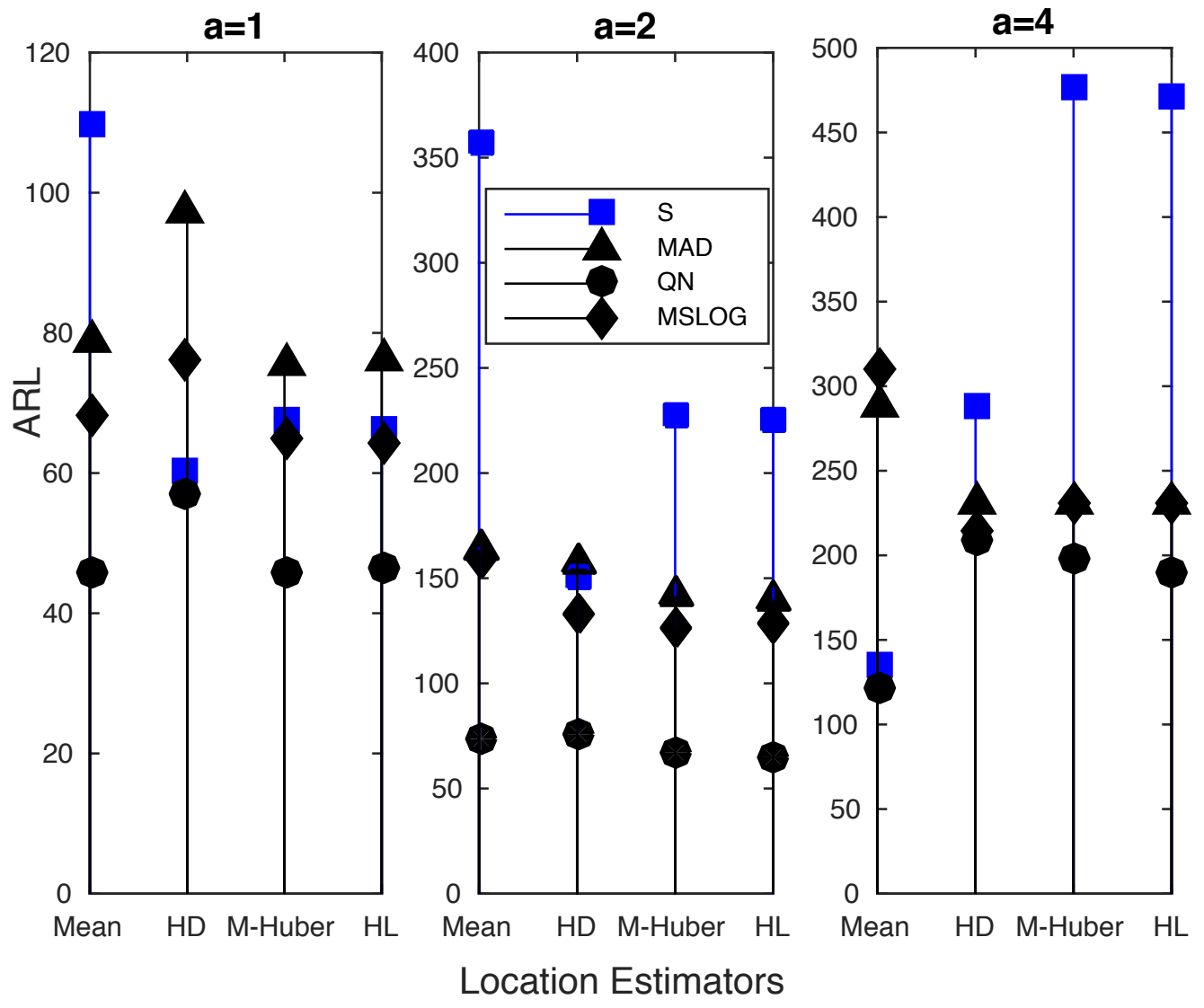

**Figure 8:** Out of control ARLs in the existence of model 2

Figure 9 shows the $ARL_0$ values for the case when model 3 is applied to Phase I sampling, and Phase II data is normally distributed with $\phi = 1$ (no disturbance in Phase II). When conventional methods are used, $ARL_0$ is 335.6 at low contamination, and severely drops down to 88.2, which is far from the design value, at high contamination. In spite of the increased contamination in Phase I sampling, $ARL_0$ value is obtained as close to the design value by using a robust location estimator. When a robust location estimator is used, particularly Harrell-Davis, sample standard deviation estimator gives the highest $ARL_0$ values, and it is followed by MSLOG. It is not surprising because model 3 is totally composed of subgroups sampled from $Z_{IC}$, however, usage of a robust location estimator is a must because 20 % of subgroups have higher standard deviation, and sample mean fails to satisfy in determination of their location. When it comes to the the case in which Phase II data is performed from disturbed process ($\phi = 1.4$), sample standard deviation is the most powerful scale estimator, notwithstanding increasing contamination (Figure 10). When sample standard deviation+sample mean is used, ARL increases to 114.9 and 470.8 at moderate and high contamination, respectively. A robust location estimator can detect the disturbance at relatively lower ARL values, here, the smallest ARL value is attained by Harrell-Davis estimator (ARL=66.2), and followed by Hodge-Leihmann and M-Huber estimators.

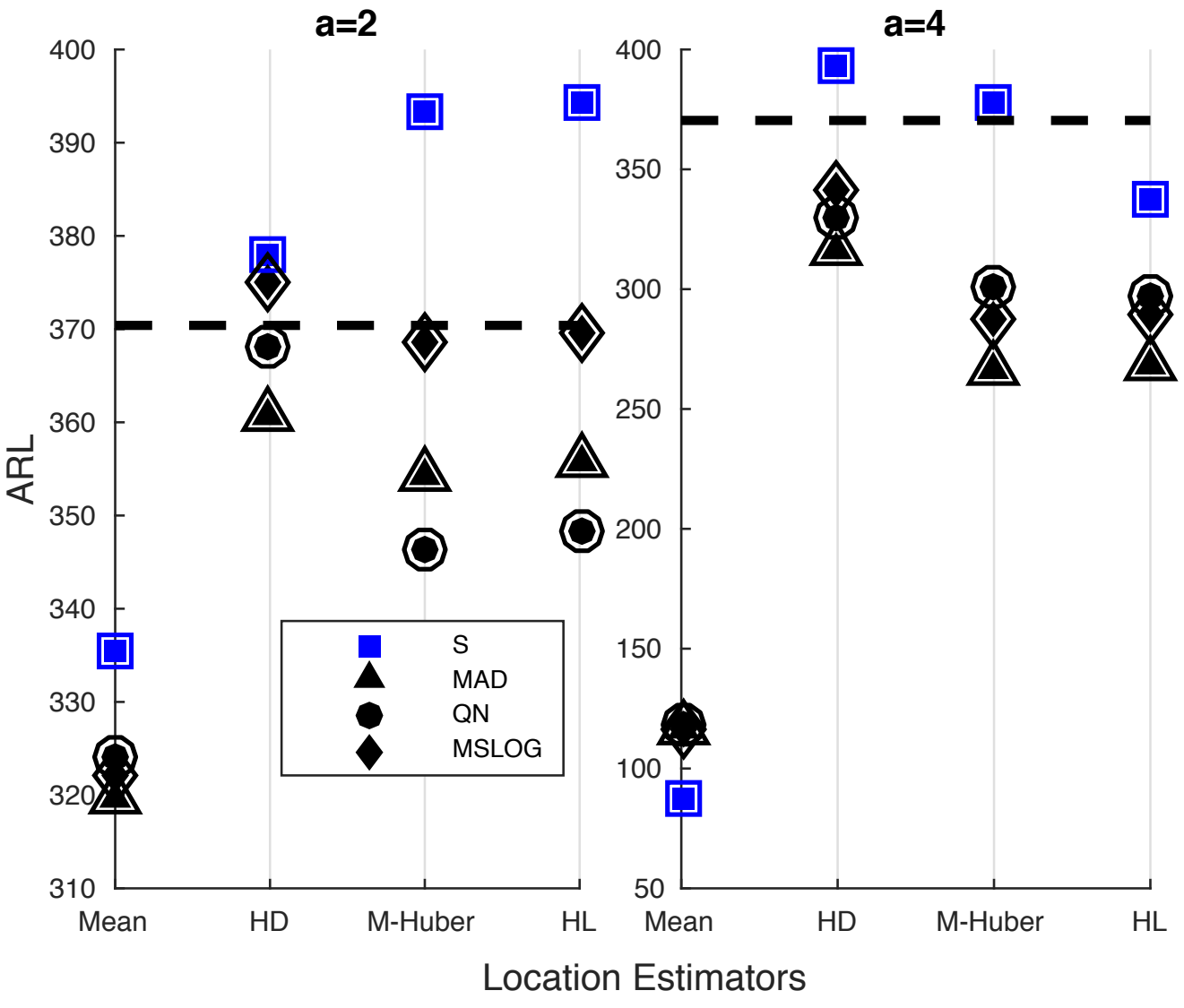

**Figure 9:** In control ARLs in the existence of model 3

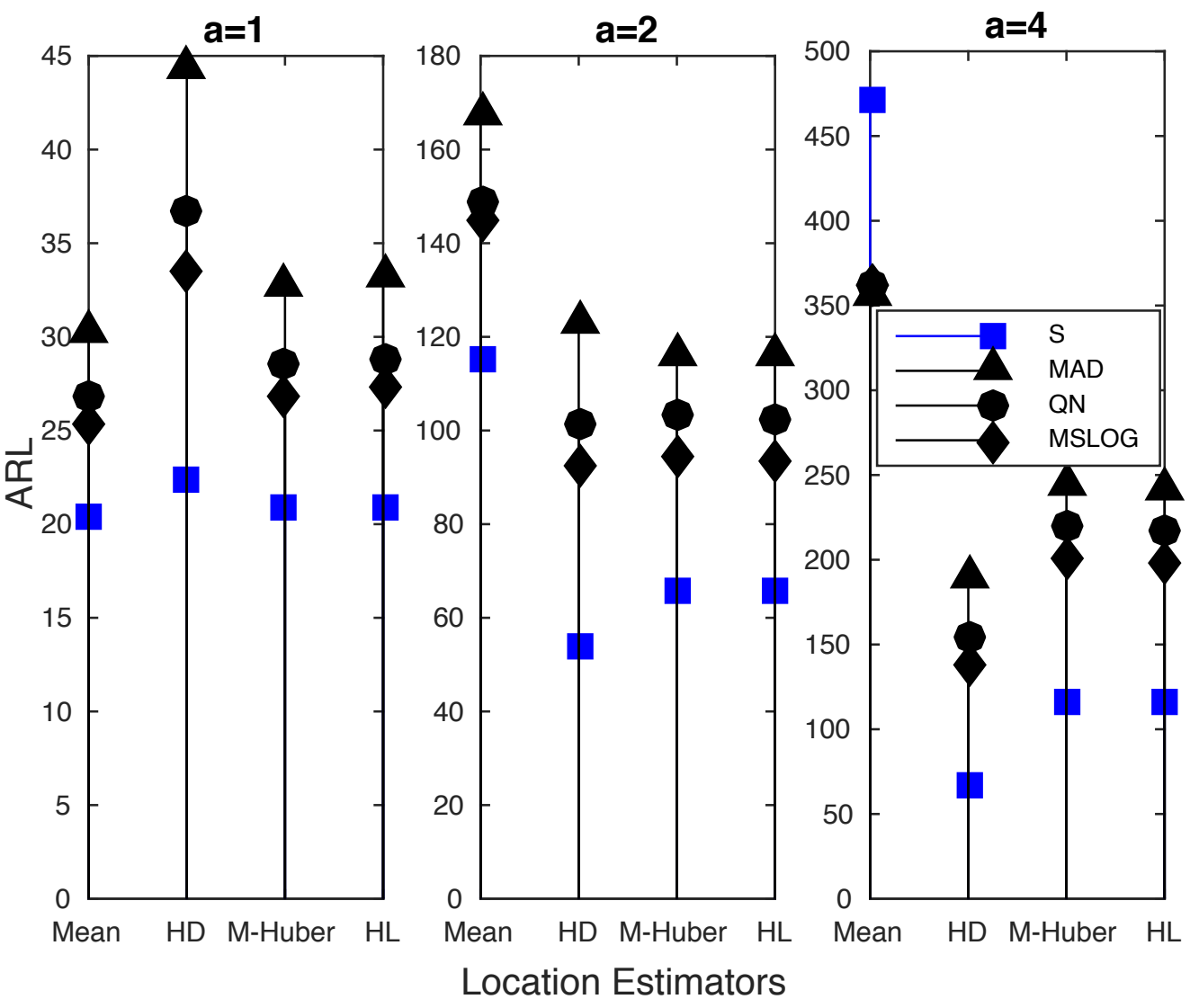

**Figure 10:** Out of control ARLs in the existence model 3

## CONCLUSION

The efficiency of conventional estimators is significantly reduced and power of Shewhart charts is undesirably low when real-world data is not normally distributed and/or outliers exist in sampling. It causes an inefficient monitoring and a serious delay in detection of inferior products. In this article, we consider several scale and location estimators for more reliable analysis. In Phase I analysis, efficiencies of proposed estimators were compared to conventional methods. When Phase I data is being solely drawn from the population of $Z_{IC}$, conventional methods estimate the population standard deviation more efficiently. However, as contamination is increased, robust estimators yield better estimation and smaller MSEs. $MAD_n$ is the most efficient scale estimator when model 1 is applied to Phase I sampling, and closely followed by MSLOG estimator. Sample standard deviation estimation is further improved using robust location estimator. The Phase I efficiencies shows their effects on Phase II sensibility and power analysis. MSLOG and $MAD_n$ is the most promising estimators when they coupled with any of the robust location estimators. The result, besides, show that usage of robust estimators more significant when contamination in Phase I sampling is still diffused but asymmetrically applied (model 2). $MAD_n$ and MSLOG scale estimators are more efficient than sample standard deviation, but QN estimator makes the best standard deviation estimations as contamination increased. $ARL_0$ for QN is found close to the design value and its power is superior compared to other statistics, the performance is followed by MSLOG and $MAD_n$ estimators. The case in which localized contamination (model 3) is applied to Phase I sampling, is dissimilar, because subgroups are sampled from normal distribution, but some of them have higher variance. Sample standard deviation has relatively higher efficiency. In Phase II analysis, sample standard deviation has the highest $ARL_0$ and lowest ARL values when it is coupled with a robust location estimator, particularly Harrell-Davis, and it is followed by MSLOG. As a result, conventional methods becomes impractical in estimation of process parameters in the existence of either diffuse-localized and/or symmetric-asymmetric contaminations. A robust scale and location estimators should be used when diffuse contamination is applied to Phase I sampling. If localized contamination is applied, robust scale estimators are still promising, but sample standard deviation is the most efficient and powerful scale estimator, when it is coupled with a robust location estimator instead of sample mean.

## NOMENCLATURE

| | |
|---|---|
| ARL | Average run length |
| ave | Sample average |
| MAD | Median absolute deviation |
| med | Sample median |
| MSLOG | Logistic M scale estimator |
| RL | Run length |
| SPC | Statistical process control |
| QN | $Q_n$ estimator of Rousseuw & Croux |
| $L_n$ | Lower control limit coefficient |
| $U_n$ | Upper control limit coefficient |
| $\alpha$ | Significance level |
| $\hat{\sigma}$ | Process standard deviation estimate |
| $\hat{\sigma}_i{}^I$ | Scale parameters of $i^{th}$ rational subgroup of Phase I sampling |
| $\hat{\sigma}_i{}^{II}$ | Scale parameters of $i^{th}$ rational subgroup of Phase II sampling |
| $\phi$ | Standard deviation multiplier |